\documentclass[aps,prl,twocolumn,superscriptaddress,preprintnumbers,floatfix]{revtex4-2}

\usepackage{float}
\usepackage{nicefrac}
\usepackage{mathtools}
\usepackage{amsfonts}
\usepackage{amssymb}
\usepackage{amsmath}
\usepackage{graphicx}
\usepackage{subfigure}
\usepackage{array}
\usepackage{dcolumn}
\usepackage[mathscr]{euscript}
\usepackage{bm}
\usepackage{latexsym}
\usepackage{longtable}
\usepackage{hyperref}
\usepackage{color}
\usepackage[capitalise]{cleveref}
\usepackage{orcidlink}
\usepackage{booktabs}
\usepackage{enumitem}

\hypersetup{
    pdfnewwindow=true,
    colorlinks=true,
    allcolors=[RGB]{31 119 180}
}

\newcommand{\bcol}{\left[ \begin{array}{c}}
\newcommand{\ecol}{\end{array} \right]}
\newcommand{\beq}{\begin{eqnarray}}
\newcommand{\eeq}{\end{eqnarray}}

\newcommand{\kev}{\ensuremath{{\mathrm{\,ke\kern -0.1em V}}}\xspace}
\newcommand{\mev}{\ensuremath{{\mathrm{\,Me\kern -0.1em V}}}\xspace}
\newcommand{\gev}{\ensuremath{{\mathrm{\,Ge\kern -0.1em V}}}\xspace}
\newcommand{\tev}{\ensuremath{{\mathrm{\,Te\kern -0.1em V}}}\xspace}

\newcommand{\mytitle}[1]{\vspace{.5cm}{\em #1.---}}

\begin{document}

\title{Neuro-dispersive extractions of light-meson resonances}
\preprint{JLAB-THY-26-4918}
\newcommand{\catania}{INFN Sezione di Catania, Catania I-95123, Italy}
\newcommand{\ceem}{Center for  Exploration  of  Energy  and  Matter, Indiana  University, Bloomington,  Indiana  47403,  USA}
\newcommand{\indiana}{Department of Physics, Indiana  University, Bloomington,  Indiana 47405,  USA}
\newcommand{\jlab}{Theory Center, Thomas  Jefferson  National  Accelerator  Facility, Newport  News,  Virginia  23606,  USA}
\newcommand{\lbnl}{Nuclear Science Division, Lawrence Berkeley National Laboratory, Berkeley, California 94720, USA}
\newcommand{\messina}{Dipartimento di Scienze Matematiche e Informatiche, Scienze Fisiche e Scienze della Terra, Universit\`a degli Studi di Messina, Messina I-98166, Italy}
\newcommand{\odu}{Department of Physics, Old Dominion University, Norfolk, Virginia 23529, USA}
\newcommand{\wm}{Department of Physics, The College of  William \& Mary, Williamsburg, Virginia 23187, USA}
\newcommand{\ucb}{Department of Physics, University of California, Berkeley, California 94720, USA}
\newcommand{\uned}{Departamento de F\'isica Interdisciplinar, Universidad Nacional de Educaci\'on a Distancia (UNED), Madrid E-28040, Spain}

\author{Wyatt~A.~Smith\,\orcidlink{0009-0001-3244-6889}}\email{wyattsmith@lbl.gov}\affiliation{\lbnl}\affiliation{\ucb}\affiliation{\wm}\affiliation{\messina}
\author{Arkaitz~Rodas\,\orcidlink{0000-0003-2702-5286}}\email{arodasbi@odu.edu}
\affiliation{\jlab}\affiliation{\odu}
\author{Marius~D.~Thomas\,\orcidlink{0009-0005-1945-594X}}\affiliation{\ucb}
\author{C\'esar~Fern\'andez-Ram\'irez\,\orcidlink{0000-0001-8979-5660}}\email{cefera@ccia.uned.es}\affiliation{\uned}
\author{Giorgio~Foti\orcidlink{0009-0000-9791-3823}}\affiliation{\messina}\affiliation{\catania}
\author{Lin~\surname{Qiu}\orcidlink{0000-0002-2683-8851}}\affiliation{\jlab}\affiliation{\odu}
\author{Adam~P.~\surname{Szczepaniak}\orcidlink{0000-0002-4156-5492}}\affiliation{\jlab}\affiliation{\ceem}\affiliation{\indiana}
\author{Alessandro~Pilloni\orcidlink{0000-0003-4257-0928}}\affiliation{\messina}\affiliation{\catania}

\collaboration{Joint Physics Analysis Center}

\date{\today}

\begin{abstract}
We present the first dispersive extraction of resonant poles from analytically continued neural networks. We use S-matrix informed neural networks (SINNs) trained to respect unitarity, analyticity, and crossing symmetry, without fixing a specific amplitude parametrization. The SINN framework controls representation dependence, enables constrained data selection, and enforces first principles. A large ensemble of networks trained on $\pi\pi$ scattering data propagates correlated uncertainties to all derived observables. We obtain robust determinations of the $\sigma/f_0(500)$, $\rho(770)$, and $f_0(980)$ poles of $\pi\pi$ scattering. Scattering lengths are determined alongside the amplitudes, while Adler zeroes emerge as predictions of the analytic structure. The results are stable against variations of the network architecture, and our approach can easily be adjusted for analysis of other reactions relevant to New Physics searches.
\end{abstract}

\maketitle

\mytitle{Introduction} The extraction of properties of unstable states from scattering data is a problem common to many areas of physics, from nuclear reactions to ultracold collisions to relativistic particle physics. In hadron spectroscopy, where the nonperturbative regime of quantum chromodynamics (QCD) lacks an analytic solution, resonances must be inferred from models of scattering data, either measured experimentally or computed from finite-volume lattice QCD spectra~\cite{Shepherd:2016dni,Briceno:2017max}. The process-independent definition of a resonance is a pole of the scattering matrix (S-matrix) in the complex energy plane. Locating these poles reliably requires the analytic continuation of a model of the scattering amplitude that is flexible enough to both fit the measurements and satisfy the S-matrix principles of unitarity, analyticity, and crossing symmetry~\cite{GoldbergerWatson:1964,Eden:1966dnq,Newton:1982,Pelaez:2015qba,ParticleDataGroup:2024cfk}.

Low-energy $\pi\pi$ scattering is one of the most pressing instances of this problem. Pions are the pseudo--Nambu-Goldstone bosons in QCD~\cite{Weinberg:1978kz,Gasser:1983yg,Gasser:1984gg,Burgess:1998ku}, making $\pi\pi$ scattering a clean probe of the chiral structure of the QCD vacuum, and of chiral perturbation theory (ChPT) at low energies. Its scalar sector contains the $\sigma/f_0(500)$ and $f_0(980)$, both states without a simple quark-model interpretation~\cite{Baru:2003qq,Jaffe:2004ph,RuizdeElvira:2010cs,Londergan:2013dza}. Their pole positions are especially sensitive to the choice of amplitude parametrization and the analytic continuation method~\cite{Caprini:2008fc,ParticleDataGroup:2024cfk}. These scalars, together with the $\rho(770)$, govern the $\pi\pi$ final-state interactions present in many precision observables, including the hadronic contribution to the muon anomalous magnetic moment, $CP$-violation studies, and heavy-meson decays~\cite{Colangelo:2018mtw,Aoyama:2020ynm,Colangelo:2017fiz,LHCb:2019jta, Bediaga:2020qxg,Garrote:2022uub,Johnson:2024omq}. A constrained $\pi\pi$ amplitude with minimal model dependence is thus important both for spectroscopy, and as a key QCD input for New Physics searches.

Two obstacles have limited this program for decades. The first is the \emph{model} problem. Parametrizations used to describe the data which exists only on the real energy axis, \textit{e.g.} Breit-Wigner forms, $K$ matrices, unitarized chiral amplitudes, conformal or Pad\'e expansions, among others~\cite{Dobado:1989qm,Dobado:1996ps,Nieves:1999bx,Oller:1997ti,Oller:1998hw,Oller:1998zr,GomezNicola:2001as,Pelaez:2004xp,GarciaMartin:2011jx,Pelaez:2026fdrpoles}, can fit well and often implement part of the required physics, but they do not enforce all S-matrix principles simultaneously. Analytic continuation away from the real-axis data is difficult to control, and the extracted pole positions remain sensitive to the assumed functional form of the amplitude, requiring dedicated systematic studies~\cite{JPAC:2018zyd,Rodas:2021tyb,JPAC:2021rxu}. More precise data alone cannot remove this representation dependence, which is largest for broad states and for resonances near thresholds~\cite{Morgan:1990ct,Hanhart:2008mx,Caprini:2005zr,Hanhart:2014ssa,Masjuan:2014psa,Caprini:2016uxy,Rodas:2023nec}. This problem manifests itself most readily in the spread of the poles obtained by different extractions present in the literature. The second is the \emph{data} problem. As pseudoscalar mesons are unstable, meson-meson scattering cannot be measured directly. The vast majority of $\pi\pi$ data were extracted from $\pi N\to\pi\pi N^\prime$ production reactions, requiring models for the production mechanism. Different analyses of the same underlying measurements yielded mutually inconsistent solutions, most datasets are old, lack modern analyses of systematic errors, and do not provide correlations~\cite{Grayer:1974cr,Hyams:1973zf,Hyams:1975mc,Protopopescu:1973sh,Estabrooks:1974vu,Cohen:1980cq,Etkin:1981sg,Kaminski:2002pe,Yndurain:2007qm,Ochs:2013gi,Perez:2015pea}. Modern kaon decay data sharpen the threshold region~\cite{Batley:2010zza,DescotesGenon:2001tn,Colangelo:2001df} but leave the intermediate-energy tensions intact. Several published amplitude extractions are further afflicted by discrete ambiguities, so the same underlying measurements admit more than one value~\cite{Barrelet:1971pw,Chung:1997qd,JointPhysicsAnalysisCenter:2023gku,Fornes:2025ldd}. Any amplitude analysis must therefore state how it fits the data, which subset of mutually incompatible experiments it considers consistent with first principles, and how those experiments were chosen.

Machine learning (ML) with physics-driven designs~\cite{Aarts:2025gyp} offers a way forward for both problems. Rather than impose a functional form on the data, one can replace a model parametrization with a neural network (NN) whose outputs are filtered through the required physics constraints. The flexibility of such a model can then allow for studies of the experimental data to produce physics-constrained data selection criteria. ML is not new to this corner of particle physics. In the extraction of parton density functions, NNs already provide flexible, statistically interpretable representations of amplitude-like objects, reducing model bias~\cite{Ball:2008by,NNPDF:2021njg}. In the S-matrix bootstrap, neural methods have mapped the space of amplitudes consistent with unitarity and crossing, but without using experimental data to select the physically realized solution~\cite{Dersy:2023job,Gumus:2024lmj}. The application to experimental data was pioneered in~\cite{Guerrieri:2024jkn}, albeit without neural methods. ML has also been applied to pole classification and simulation-based inference in restricted settings~\cite{Sombillo:2021yxe,Ng:2021ibr,Sadasivan:2025kjj}. None of these applications have yet taken the leap to directly represent the S-matrix, confronting experimental data directly with first principles.

In this Letter, we use S-matrix informed neural networks (SINNs)~\cite{companion} to dispersively extract light-meson resonances from pion scattering data. A SINN is a direct neural representation of the scattering amplitude, constructed and trained to satisfy all relevant S-matrix and kinematic constraints. Uncertainties are faithfully propagated through independent networks trained on a parametric bootstrap of the data so that each replica individually satisfies first principles. 

\mytitle{Neuro-dispersive amplitudes}
The experimental observables available for analysis are the phase shifts $\delta_\ell^I(s)$ and inelasticities $\eta_\ell^I(s)$ of definite isospin $I$ and angular momentum $\ell$, which build the $\pi\pi$ partial waves as
\begin{align}
t_\ell^I(s)=\frac{\sqrt{s}}{2q}\frac{1}{2i}
\left(\eta_\ell^I(s)e^{2i\delta_\ell^I(s)}-1\right),
\label{eq:partial-wave}
\end{align}
with $s$ the center-of-mass energy squared and \mbox{$q=\sqrt{s-4m_\pi^2}/2$} the corresponding momentum, and \mbox{$\eta_\ell^I= 1$} in the elastic region.

A SINN replaces the explicit parametrization of these observables by a neural representation evaluated on the real axis. Its outputs pass through specialized activation functions which enforce hard constraints so every network satisfies correct threshold kinematic behavior, and the unitarity bound $0\leq\eta_\ell^I\leq1$ exactly. We describe seven partial waves, in spectroscopic notation $S_0$, $P_1$, $S_2$, $D_0$, $D_2$, $F_1$, and $G_2$.

Hard kinematic constraints and unitarity are imposed exactly on the NN by construction, while analyticity and crossing symmetry are imposed through truncated Roy equations, dispersion relations specialized to the $\pi\pi$ system.

The Roy equations play a dual role here; they are imposed on the network through the training objective, and they are also the method through which we  analytically continue the SINN-derived amplitude. In compact form, the Roy-reconstructed partial wave amplitude for $s$ inside the domain of validity is,
\begin{align}
\tilde t_\ell^I(s)
&= \mathrm{ST}_\ell^I(s)+\mathrm{DT}_\ell^I(s) \nonumber\\
&\quad+\sum_{\ell^\prime I^\prime} \int_{4m_\pi^2}^{s_h}\!\!ds^\prime\,
K_{\ell\ell^\prime}^{II^\prime}(s,s^\prime)\,
\mathrm{Im}\,t_{\ell^\prime}^{I^\prime}(s^\prime).
\label{eq:roy-compact}
\end{align}
The subtraction terms in $\mathrm{ST}$ are first degree polynomials fixed by the $S$-wave scattering lengths, the kernels $K_{\ell\ell^\prime}^{II^\prime}$ encode crossing symmetry, and $\mathrm{DT}$ contains the high-energy driving terms corresponding to the completion of the integral evaluated with a fixed Regge representation above the matching point $\sqrt{s_h} = 1.5~\mathrm{GeV}$. The $\tilde{t}_\ell^I(s)$ denotes the Roy-reconstructed partial wave, while $t_{\ell^\prime}^{I^\prime}$ denotes the SINN-derived amplitude. The Roy equations are imposed only for the dominant $S_0$, $P_1$, and $S_2$ waves; the remaining four enter the dispersive integrals as input. SINN training seeks to minimize the difference between the Roy-reconstructed and neural representations of the partial waves, while matching the data.

Uncertainties come from Gaussian replicas~\cite{Ball:2008by}: every measurement is shifted within its quoted uncertainty and an independent SINN is retrained. Covariance matrices are unavailable for most historical inputs, so the replica shifts are independent. Because every replica passes through the same kinematical constraints and coupled Roy system, the resulting uncertainty bands are empirical distributions of fully constrained amplitudes, and any derived observable can be evaluated replica by replica and aggregated across the ensemble. Every replica within a SINN ensemble respects crossing symmetry, unlike traditional analyses. SINNs propagate correlations between partial waves throughout the entire analysis, whereas independently fitting partial waves accrues errors by neglecting these correlations.

\begin{figure}
\centering
\includegraphics[width=\linewidth]{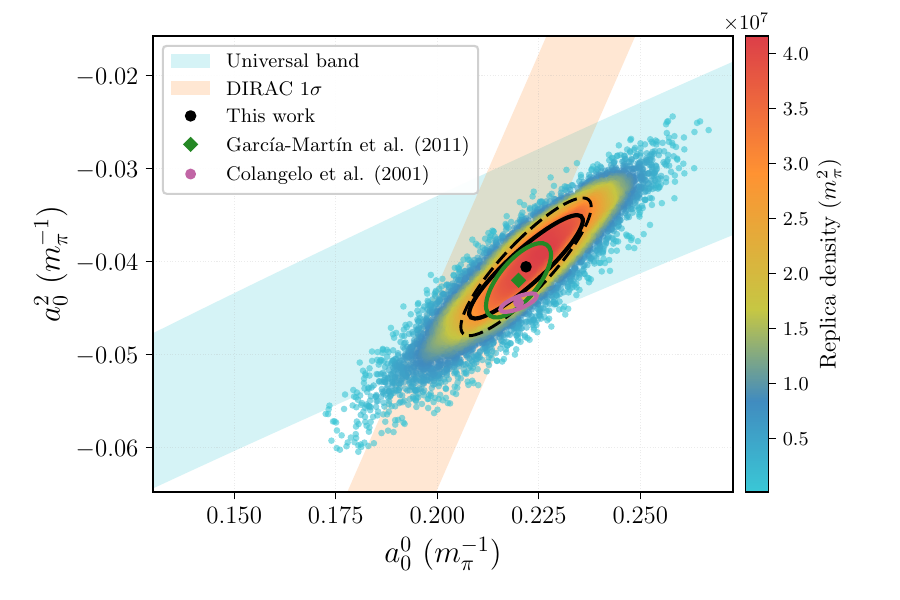} 
\caption{Joint distribution of scattering lengths $a_0^0$ and $a_0^2$ across the production ensemble, colored by local density. The solid ellipse has half-extents equal to the central $68\%$ interval of~\cref{tab:scattering_lengths}. The dashed ellipse adds the spread of the architecture sweep of Appendix~\ref{app:arch} in quadrature, and estimates a small systematic shift in uncertainties, with no shift in central value. Reference determinations from Refs.~\cite{Colangelo:2001df, GarciaMartin:2011cn} are shown with their error ellipses.}
\label{fig:scatt}
\end{figure}

\begin{table}
\caption{Scattering lengths in units of $m_\pi^{-1}$. The uncertainty on this work is the central $68\%$ interval over the production ensemble.}
\label{tab:scattering_lengths}
\centering
\small
\begin{tabular}{c c c c}
\toprule
 & This work & Ref.~\cite{Colangelo:2001df} & Ref.~\cite{GarciaMartin:2011cn} \\
\midrule
$a_0^0$ & $0.222^{+0.014}_{-0.014}$ & $0.220 \pm 0.005$ & $0.220 \pm 0.008$ \\[2pt]
$a_0^2$ & $-0.041^{+0.005}_{-0.006}$ & $-0.0444\pm0.0010$ & $-0.042\pm 0.004$ \\
\bottomrule
\end{tabular}
\end{table}

In performing analytic continuation for extraction of off-axis observables, the SINN is never evaluated for complex energies. Complex energies $s$ only appear in the Roy subtraction terms, kernels, and driving term, while the SINN is evaluated only for real values of $s^\prime$ under the integral. The pole calculation is therefore a dispersive continuation of constrained real-axis input, and repeating it replica by replica propagates the correlated ensemble uncertainties to the pole positions.

Constrained uncertainty bands from data replicas additionally allow for constrained data selection~\cite{NNPDF:2021njg,Ward01031963,Murtagh2014,KohLiang2017,Ilyas2022}. We perform a novel data selection procedure \emph{spectral response clustering}: each candidate experiment is upweighted inside independent SINN ensembles, and the set of all such ensembles is compared to a uniformly weighted ensemble to study the statistical compatibilities of each experiment within the larger class of constrained amplitudes. The pattern of responses is spectrally decomposed and clustered to identify mutually compatible groups, and persistently incompatible groups are removed iteratively. Such an analysis is only meaningful because the SINN can follow the pull of one experiment without sacrificing satisfaction of the physics constraints. A rigid parametrization would conflate experimental tension with model rigidity. Spectral response clustering is detailed in Ref.~\cite{companion}; it leaves a reproducible, physics-constrained selection of 18 experiments and 621 measurements in $0.28\leq\sqrt{s}\leq1.5~\mathrm{GeV}$. Our production ensemble  presented below contains $10{,}000$ independently trained networks on this selected data set.

\mytitle{Results}
The $S$-wave scattering lengths are extracted from the same constrained ensemble replica by replica. \Cref{tab:scattering_lengths,fig:scatt} show the joint distribution of the scattering lengths, where $a_0^0$ and $a_0^2$ track each other with very high correlation. They agree with previous dispersive analyses, Colangelo et al.~\cite{Colangelo:2001df} and Garc\'ia-Mart\'in et al.~\cite{GarciaMartin:2011cn}. Our uncertainties are larger, which is expected as we probe a larger space of constrained solutions. The much tighter interval quoted by Colangelo et al. follows from matching Roy solutions onto output from two-loop ChPT instead of determining them from the scattering data. García-Martín et al. impose the chiral Adler zeroes directly into their initial model fit. Both impose additional model constraints we do not include. The SINN amplitude accurately reproduces all experimental data used, including DIRAC's most recent measurement~\cite{Adeva:2011tc}. Our result also lies within the so-called ``universal band''~\cite{Ananthanarayan:2000ht}, considered as the band in the scattering length plane where solutions to $\pi \pi$ Roy equations can occur.

\begin{figure}[!ht]
\centering
\includegraphics[width=\linewidth]{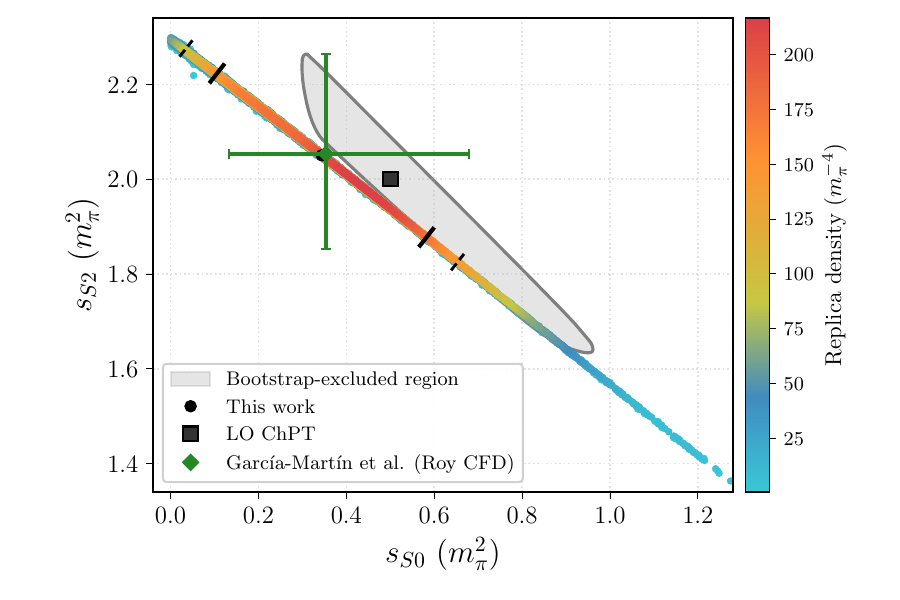}
\caption{Predicted Adler zero positions determined by Roy equations. The two zeros are almost perfectly anticorrelated. The central black circle describes the median and the filled bars describe the standard deviation interval of the production model, and the dashed bars show the statistical plus systematic uncertainty, taken from the architecture sweep of Appendix~\ref{app:arch}. The result is compared with Ref.~\cite{GarciaMartin:2011cn} (green), presented here as uncorrelated, and the exclusion region obtained with bootstrap for fixed $\rho$ mass~\cite{Guerrieri:2018uew}.}
\label{fig:adler}
\end{figure}

\begin{figure*}[!ht]
\centering
\includegraphics[width=\linewidth]{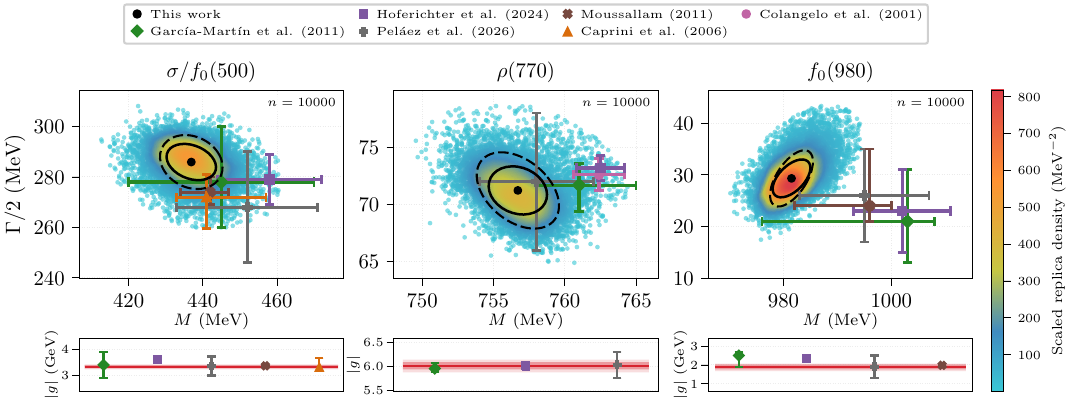}
\caption{Pole positions in the complex $\sqrt{s}$ plane on the second ($\pi\pi$) Riemann sheet, together with coupling comparisons. Each point is one replica of the production ensemble, colored by local density. The solid ellipse has half-extents equal to the central $68\%$ interval of~\cref{tab:poles}, and the dashed ellipse adds the spread of the architecture sweep of Appendix~\ref{app:arch} as an estimate for systematic uncertainty on the production interval; the couplings show the same two bands. Literature values are shown as markers with uncertainties~\cite{GarciaMartin:2011jx,Colangelo:2001df,Hoferichter:2023mgy,Pelaez:2026fdrpoles,Moussallam:2011zg,Caprini:2005zr}.}
\label{fig:poles}
\end{figure*}

Analytic continuation of the amplitude also allows us to analyze the subthreshold region. Chiral symmetry requires the $S_0$ and $S_2$ amplitudes to vanish at the Adler points, which for leading order ChPT appear at \mbox{$s_{S_0}=m_\pi^2/2$} and \mbox{$s_{S_2}=2m_\pi^2$}~\cite{Adler:1964um,Gasser:1983yg,Gasser:1984gg}. Dispersively, we find the $S_0$ zero at \mbox{$s=0.3^{+0.3}_{-0.2}\,m_\pi^2$} and the $S_2$ zero at \mbox{$s=2.1^{+0.2}_{-0.2}\,m_\pi^2$}, shown in \cref{fig:adler}. As no chiral input enters our analysis, the extracted zeros are genuine predictions from our SINN ensemble. In the figure, our results are compared to those of~\cite{GarciaMartin:2011cn}, and the bootstrap analysis of~\cite{Guerrieri:2018uew}. Our results lie on a line, almost parallel and extremely close to the boundary of the excluded region from the bootstrap analysis. Since theories at this boundary are uniquely defined, this fact suggests that physical QCD may play a special role among generic theories with the same chiral symmetry breaking pattern. Additional sum rules might further constrain the position of Adler zeros.
We remark that roughly $20\%$ of all replicas did not produce an $S_0$ zero. We find an accumulation of zeroes near $m_\pi^2 = 0$, which risks overlap with the left hand cut. We also observe a strong correlation between larger values of the \(a_0^0\) scattering length and \(S_0\) Adler zeros approaching \(s=0\). We do not consider potential Adler zeroes which would have produced this unphysical behavior. 

The pole positions and couplings also follow from analytic continuation. The conjugate lower-half-plane poles are reported as \mbox{$M-i\Gamma/2$}, and the couplings are obtained from their residues in the standard normalization of the literature~\cite{Pelaez:2020gnd}. The resulting ensemble distributions are shown in~\cref{fig:poles}, and numerical values are summarized in~\cref{tab:poles}.

Resonance pole extraction through a SINN is extremely stable: all three resonance poles were reliably and independently extracted in every single network in the ensemble. Our $\sigma/f_0(500)$ pole agrees with every determination in~\cref{tab:poles}, and all three widths and couplings are compatible throughout. The $\rho(770)$ and $f_0(980)$ masses agree with the most recent dispersive determination of Ref.~\cite{Pelaez:2026fdrpoles}. We report the decay width of the $f_0(980)$ as consistent with traditional analyses, but find a shifted mass, likely caused by the absence of a $K\bar K$ cusp in the smooth SINN representation of the amplitude; the available data at the location of the sharp rise in the $S_0$ phase shift is not dense enough to cleanly resolve such a cusp, and we did not enforce the discontinuity in the derivative by hand. Hence our SINNs provide a smooth interpolation over the region, while the literature enforces this cusp. Remarkably, all reported pole positions, scattering lengths, and Adler zeros are stable against variations in architecture and hyperparameter choices, indicating negligible systematic uncertainties. This architecture independence reflects the fact that the SINN fully explores the functional space allowed by the physical constraints. We show the specifics of the analytic continuation, and several validation and robustness studies for all extracted quantities in the End Matter.

\begin{table*}[!ht]
\caption{Pole positions and couplings compared with representative dispersive and bootstrap determinations. Pole positions are \mbox{$\sqrt{s_{\rm pole}}=M-i\Gamma/2$} in $\mathrm{MeV}$ on the second ($\pi\pi$) Riemann sheet. Scalar couplings are quoted in $\mathrm{GeV}$, while $|g_{\rho\pi\pi}|$ is dimensionless in the standard normalization used in the pole analysis. The uncertainty on this work is the central 68\% interval over the $10{,}000$-replica production ensemble. The Moussallam couplings are converted to the normalization of Eq.~\eqref{eq:endmatter_coupling}.}
\label{tab:poles}

\centering
\footnotesize
\setlength{\tabcolsep}{3.5pt}
\begin{tabular}{l c c c c c c}
\toprule
Reference
& $\sqrt{s_{\sigma/f_0(500)}}$  & $|g_{\sigma\pi\pi}|$ 
& $\sqrt{s_{\rho(770)}}$  & $|g_{\rho\pi\pi}|$
& $\sqrt{s_{f_0(980)}}$  & $|g_{f_0\pi\pi}|$  \\
\midrule
{\bf This work}
  & $437^{+7}_{-6}-i\,286^{+6}_{-8}$
  & $3.33^{+0.06}_{-0.06}$
  & $757^{+2}_{-2}-i\,71^{+2}_{-2}$
  & $6.00^{+0.08}_{-0.07}$
  & $981^{+4}_{-3}-i\,29^{+4}_{-3}$
  & $1.87^{+0.15}_{-0.11}$ \\ [3pt]
  \hline \\ [-5pt]
Garc\'ia-Mart\'in et al.~\cite{GarciaMartin:2011jx}
  & $445\pm25-i\,278^{+22}_{-18}$
  & $3.4\pm0.5$
  & $761^{+4}_{-3}-i\,71.7^{+1.9}_{-2.3}$
  & $5.95^{+0.12}_{-0.08}$
  & $1003^{+5}_{-27}-i\,21^{+10}_{-8}$
  & $2.5^{+0.2}_{-0.6}$ \\[3pt]
Colangelo et al.~\cite{Colangelo:2001df}
  & $-$
  & $-$
  & $762.4\pm1.8-i\,72.6\pm1.4$
  & $-$
  & $-$
  & $-$ \\[3pt]
Hoferichter et al.~\cite{Hoferichter:2023mgy}
  & $458\pm14-i\,279\pm10$
  & $3.61\pm0.13$
  & $762.5\pm1.7-i\,73.2\pm1.1$
  & $6.01\pm0.08$
  & $1002\pm9-i\,23\pm8$
  & $2.33\pm0.18$ \\[3pt]
Pel\'aez et al.~\cite{Pelaez:2026fdrpoles}
  & $452\pm19-i\,268\pm22$
  & $3.36\pm0.36$
  & $758\pm4-i\,72\pm6$
  & $6.025\pm0.275$
  & $995\pm12-i\,26\pm9$
  & $1.9\pm0.6$ \\[3pt]
Moussallam~\cite{Moussallam:2011zg}
  & $442^{+5}_{-8}-i\,274^{+6}_{-5}$
  & $3.37$
  & $-$
  & $-$
  & $996^{+4}_{-14}-i\,24^{+11}_{-3}$
  & $1.98$ \\[3pt]
Caprini et al.~\cite{Caprini:2005zr}
  & $441^{+16}_{-8}-i\,272^{+9}_{-13}$
  & $3.31^{+0.35}_{-0.15}$
  & $-$
  & $-$
  & $-$
  & $-$ \\
\midrule
Guerrieri et al.~\cite{Guerrieri:2024jkn}  
  & $413\pm3-i\,289\pm3$
  & $-$
  & $765.6\pm0.7-i\,74.3\pm0.3$
  & $-$
  & $991\pm6-i\,36\pm3$
  & $-$ \\
\bottomrule
\end{tabular}
\end{table*}

\mytitle{Summary}
The SINN supplies a flexible real-axis representation that satisfies kinematic constraints exactly and first principles numerically. The resulting ensembles of fully constrained amplitude replicas allow for constrained uncertainty propagation to all derived observables, and enables constrained data-selection~\cite{companion}. SINN-based analysis remedies both the model and data problems associated with amplitude analysis. It provides control over the epistemic and stochastic uncertainties that propagate into every derived quantity. SINNs provide robust scattering length and subthreshold Adler zero estimates, and resonance pole parameters for the $\sigma/f_0(500)$, $\rho(770)$, and $f_0(980)$, appearing in $\pi\pi$ scattering. We find negligible systematic uncertainties in all extracted quantities.

The same framework can be applied to other systems where data tensions and model dependence limit the precision of amplitude analyses, including $\pi K$ and $\pi N$ scattering, where the hadronic inputs feed directly into neutrino-nucleus cross-section models, and hadronic form factors entering the hadronic vacuum polarization contribution to $(g-2)_\mu$.

\mytitle{Data availability}
The production SINN ensemble replicas, medians, and confidence intervals used to obtain our results are provided along with an example notebook at Ref.~\cite{wyatt_a_smith_2026_22016314}.

\mytitle{Acknowledgments}
We thank Yaohang Li for his careful reading of the manuscript and suggestions. We thank Emanuele Roberto Nocera and Amedeo Chiefa for helpful discussions on the NNPDF framework. We also thank Marco Filippini and Antonino Fulci for their collaboration in the early stages of this work. This material is based upon work supported by the U.S. Department of Energy, Office of Science, Office of Nuclear Physics under Contract No. 89243126CSC000213, by U.S.~Department of Energy Grant
Nos.~\mbox{DE-FG02-87ER40365}, and \mbox{DE-SC0011090}, and it contributes to the aims of the
U.S.~Department of Energy \mbox{ExoHad} Topical Collaboration, contract \mbox{DE-SC0023598}. ARB acknowledges support by the National Science Foundation (NSF) under Grant No. PHY-2610011. AP and WAS acknowledge support of the GeV-AI project (CUP I57G21000110007), in the context of the ICSC Spoke 2 Open Calls, project funded by European Union -- NextGenerationEU -- and National Recovery and Resilience Plan (NRRP) -- Mission 4 Component 2, and support under the program
``Progetti di Rilevante Interesse Nazionale''
\mbox{(PRIN~2022)}, published on \mbox{2.2.2022} by the Italian Ministry of University and Research (MUR),
Project Title ``The X(3872) files'' -- \mbox{CUP~J53C24002600006} -- Grant Assignment Decree
\mbox{No.~20429} adopted on \mbox{6.11.2024} by the Italian Ministry of University and Research (MUR). This work was supported by the Research Computing clusters at Old Dominion University. The Wahab cluster at Old Dominion University is supported in part by National Science Foundation's grant CNS-1828593.
The authors acknowledge William \& Mary Research Computing for providing computational resources that have contributed to the results reported within this paper.

\bibliographystyle{apsrev4-2}
\bibliography{bibliography}

@article{Aarts:2025gyp,
    author = "Aarts, Gert and Fukushima, Kenji and Hatsuda, Tetsuo and Ipp, Andreas and Shi, Shuzhe and Wang, Lingxiao and Zhou, Kai",
    title = "{Physics-driven learning for inverse problems in quantum chromodynamics}",
    eprint = "2501.05580",
    archiveprefix = "arXiv",
    primaryclass = "hep-lat",
    reportnumber = "RIKEN-iTHEMS-Report-25",
    doi = "10.1038/s42254-024-00798-x",
    journal = "Nature Rev. Phys.",
    volume = "7",
    number = "3",
    pages = "154--163",
    year = "2025"
}

@article{Adeva:2011tc,
    author = "Adeva, B. and others",
    title = "{Determination of $\pi\pi$ scattering lengths from measurement of $\pi^+\pi^-$ atom lifetime}",
    eprint = "1109.0569",
    archiveprefix = "arXiv",
    primaryclass = "hep-ex",
    reportnumber = "CERN-PH-EP-2011-028, FERMILAB-PUB-10-713-E, DIRAC-PUB-2011-01",
    doi = "10.1016/j.physletb.2011.08.074",
    journal = "Phys. Lett. B",
    volume = "704",
    pages = "24--29",
    year = "2011"
}

@article{Adler:1964um,
    author = "Adler, Stephen L.",
    title = "{Consistency conditions on the strong interactions implied by a partially conserved axial vector current}",
    doi = "10.1103/PhysRev.137.B1022",
    journal = "Phys. Rev.",
    volume = "137",
    pages = "B1022--B1033",
    year = "1965"
}

@article{Ananthanarayan:2000ht,
    author = "Ananthanarayan, B. and Colangelo, G. and Gasser, J. and Leutwyler, H.",
    title = "{Roy equation analysis of $\pi \pi$ scattering}",
    eprint = "hep-ph/0005297",
    archiveprefix = "arXiv",
    reportnumber = "IISC-CTS-12-99, ZU-TH-10-00, BUTP-99-33",
    doi = "10.1016/S0370-1573(01)00009-6",
    journal = "Phys. Rept.",
    volume = "353",
    pages = "207--279",
    year = "2001"
}

@article{Aoyama:2020ynm,
    author = "Aoyama, T. and others",
    title = "{The anomalous magnetic moment of the muon in the Standard Model}",
    eprint = "2006.04822",
    archiveprefix = "arXiv",
    primaryclass = "hep-ph",
    reportnumber = "FERMILAB-PUB-20-207-T, INT-PUB-20-021, KEK Preprint 2020-5, MITP/20-028, KEK Preprint 2020-5, MITP/20-028, CERN-TH-2020-075, IFT-UAM/CSIC-20-74, LMU-ASC 18/20, LTH 1234, LU TP 20-20, LTH 1234, LU TP 20-20, MAN/HEP/2020/003, PSI-PR-20-06, UWThPh 2020-14, ZU-TH 18/20",
    doi = "10.1016/j.physrep.2020.07.006",
    journal = "Phys. Rept.",
    volume = "887",
    pages = "1--166",
    year = "2020"
}

@article{Ball:2008by,
    author = "Ball, Richard D. and Del Debbio, Luigi and Forte, Stefano and Guffanti, Alberto and Latorre, Jose I. and Piccione, Andrea and Rojo, Juan and Ubiali, Maria",
    collaboration = "NNPDF",
    title = "{A Determination of parton distributions with faithful uncertainty estimation}",
    eprint = "0808.1231",
    archiveprefix = "arXiv",
    primaryclass = "hep-ph",
    reportnumber = "EDINBURGH-2008-25, IFUM-923-FT, FREIBURG-2008-08",
    doi = "10.1016/j.nuclphysb.2008.09.037",
    journal = "Nucl. Phys. B",
    volume = "809",
    pages = "1--63",
    year = "2009",
    note = "[Erratum: Nucl.Phys.B 816, 293 (2009)]"
}

@article{Barrelet:1971pw,
    author = "Barrelet, E.",
    title = "{A new point of view in the analysis of two-body reactions}",
    doi = "10.1007/BF02732655",
    journal = "Nuovo Cim. A",
    volume = "8",
    pages = "331--371",
    year = "1972"
}

@article{Baru:2003qq,
    author = "Baru, V. and Haidenbauer, J. and Hanhart, C. and Kalashnikova, Yu. and Kudryavtsev, Alexander Evgenyevich",
    title = "{Evidence that the $a_0(980)$ and $f_0(980)$ are not elementary particles}",
    eprint = "hep-ph/0308129",
    archiveprefix = "arXiv",
    reportnumber = "FZJ-IKP-TH-2003-13",
    doi = "10.1016/j.physletb.2004.01.088",
    journal = "Phys. Lett. B",
    volume = "586",
    pages = "53--61",
    year = "2004"
}

@article{Batley:2010zza,
    author = "Batley, J. R. and others",
    collaboration = "NA48/2",
    title = "{Precise tests of low energy QCD from $K_{e4}$ decay properties}",
    doi = "10.1140/epjc/s10052-010-1480-6",
    journal = "Eur. Phys. J. C",
    volume = "70",
    pages = "635--657",
    year = "2010"
}

@article{Bediaga:2020qxg,
    author = "Bediaga, Ignacio and G{\"o}bel, Carla",
    title = "{Direct $CP$ violation in beauty and charm hadron decays}",
    eprint = "2009.07037",
    archiveprefix = "arXiv",
    primaryclass = "hep-ex",
    doi = "10.1016/j.ppnp.2020.103808",
    journal = "Prog. Part. Nucl. Phys.",
    volume = "114",
    pages = "103808",
    year = "2020"
}

@article{Briceno:2017max,
    author = "Briceno, Raul A. and Dudek, Jozef J. and Young, Ross D.",
    title = "{Scattering processes and resonances from lattice QCD}",
    eprint = "1706.06223",
    archiveprefix = "arXiv",
    primaryclass = "hep-lat",
    reportnumber = "JLAB-THY-17-2495, ADP-17-28-T1034",
    doi = "10.1103/RevModPhys.90.025001",
    journal = "Rev. Mod. Phys.",
    volume = "90",
    number = "2",
    pages = "025001",
    year = "2018"
}

@article{Burgess:1998ku,
    author = "Burgess, C. P.",
    title = "{Goldstone and pseudoGoldstone bosons in nuclear, particle and condensed matter physics}",
    eprint = "hep-th/9808176",
    archiveprefix = "arXiv",
    reportnumber = "MCGILL-98-25",
    doi = "10.1016/S0370-1573(99)00111-8",
    journal = "Phys. Rept.",
    volume = "330",
    pages = "193--261",
    year = "2000"
}

@article{Caprini:2005zr,
    author = "Caprini, Irinel and Colangelo, Gilberto and Leutwyler, Heinrich",
    title = "{Mass and width of the lowest resonance in QCD}",
    eprint = "hep-ph/0512364",
    archiveprefix = "arXiv",
    doi = "10.1103/PhysRevLett.96.132001",
    journal = "Phys. Rev. Lett.",
    volume = "96",
    pages = "132001",
    year = "2006"
}

@article{Caprini:2008fc,
    author = "Caprini, Irinel",
    title = "{Finding the $\sigma$ pole by analytic extrapolation of $\pi \pi$ scattering data}",
    eprint = "0804.3504",
    archiveprefix = "arXiv",
    primaryclass = "hep-ph",
    doi = "10.1103/PhysRevD.77.114019",
    journal = "Phys. Rev. D",
    volume = "77",
    pages = "114019",
    year = "2008"
}

@article{Caprini:2016uxy,
    author = "Caprini, Irinel and Masjuan, Pere and Ruiz de Elvira, Jacobo and Sanz-Cillero, Juan Jos{\'e}",
    title = "{Uncertainty estimates of the $\sigma$ pole determination by Pad{\'e} approximants}",
    eprint = "1602.02062",
    archiveprefix = "arXiv",
    primaryclass = "hep-ph",
    doi = "10.1103/PhysRevD.93.076004",
    journal = "Phys. Rev. D",
    volume = "93",
    number = "7",
    pages = "076004",
    year = "2016"
}

@article{Chung:1997qd,
    author = "Chung, S. U.",
    title = "{Techniques of amplitude analysis for two pseudoscalar systems}",
    reportnumber = "BNL-QGS-97-041",
    doi = "10.1103/PhysRevD.56.7299",
    journal = "Phys. Rev. D",
    volume = "56",
    pages = "7299--7316",
    year = "1997"
}

@article{Cohen:1980cq,
    author = "Cohen, Daniel H. and Ayres, D. S. and Diebold, R. and Kramer, S. L. and Pawlicki, A. J. and Wicklund, A. B.",
    title = "{Amplitude Analysis of the $K^- K^+$ System Produced in the Reactions $\pi^- p \to K^- K^+ n$ and $\pi^+ n \to K^- K^+ p$ at 6 GeV$/c$}",
    reportnumber = "ANL-HEP-PR-80-23",
    doi = "10.1103/PhysRevD.22.2595",
    journal = "Phys. Rev. D",
    volume = "22",
    pages = "2595",
    year = "1980"
}

@article{Colangelo:2001df,
    author = "Colangelo, G. and Gasser, J. and Leutwyler, H.",
    title = "{$\pi \pi$ scattering}",
    eprint = "hep-ph/0103088",
    archiveprefix = "arXiv",
    reportnumber = "ZU-TH-3-01, BUTP-01-1",
    doi = "10.1016/S0550-3213(01)00147-X",
    journal = "Nucl. Phys. B",
    volume = "603",
    pages = "125--179",
    year = "2001"
}

@article{Colangelo:2017fiz,
    author = "Colangelo, Gilberto and Hoferichter, Martin and Procura, Massimiliano and Stoffer, Peter",
    title = "{Dispersion relation for hadronic light-by-light scattering: two-pion contributions}",
    eprint = "1702.07347",
    archiveprefix = "arXiv",
    primaryclass = "hep-ph",
    reportnumber = "INT-PUB-17-009, CERN-TH-2017-041, NSF-KITP-17-036",
    doi = "10.1007/JHEP04(2017)161",
    journal = "JHEP",
    volume = "04",
    pages = "161",
    year = "2017"
}

@article{Colangelo:2018mtw,
    author = "Colangelo, Gilberto and Hoferichter, Martin and Stoffer, Peter",
    title = "{Two-pion contribution to hadronic vacuum polarization}",
    eprint = "1810.00007",
    archiveprefix = "arXiv",
    primaryclass = "hep-ph",
    reportnumber = "INT-PUB-18-048",
    doi = "10.1007/JHEP02(2019)006",
    journal = "JHEP",
    volume = "02",
    pages = "006",
    year = "2019"
}

@article{DelDebbio:2021whr,
    author = "Del Debbio, Luigi and Giani, Tommaso and Wilson, Michael",
    title = "{Bayesian approach to inverse problems: an application to NNPDF closure testing}",
    eprint = "2111.05787",
    archiveprefix = "arXiv",
    primaryclass = "hep-ph",
    doi = "10.1140/epjc/s10052-022-10297-x",
    journal = "Eur. Phys. J. C",
    volume = "82",
    number = "4",
    pages = "330",
    year = "2022"
}

@article{Dersy:2023job,
    author = "Dersy, Aur{\'e}lien and Schwartz, Matthew D. and Zhiboedov, Alexander",
    title = "{Reconstructing S-matrix Phases with Machine Learning}",
    eprint = "2308.09451",
    archiveprefix = "arXiv",
    primaryclass = "hep-th",
    reportnumber = "CERN-TH-2023-161",
    doi = "10.1007/JHEP05(2024)200",
    journal = "JHEP",
    volume = "05",
    pages = "200",
    year = "2024"
}

@article{DescotesGenon:2001tn,
    author = "Descotes-Genon, S. and Fuchs, N. H. and Girlanda, L. and Stern, J.",
    title = "{Analysis and interpretation of new low-energy $\pi \pi$ scattering data}",
    eprint = "hep-ph/0112088",
    archiveprefix = "arXiv",
    reportnumber = "SHEP-01-32, DFPD-01-TH-33, IPNO-DR-01-024",
    doi = "10.1007/s10052-002-0965-3",
    journal = "Eur. Phys. J. C",
    volume = "24",
    pages = "469--483",
    year = "2002"
}

@article{Dobado:1989qm,
    author = "Dobado, A. and Herrero, Maria J. and Truong, Tran N.",
    title = "{Unitarized Chiral Perturbation Theory for Elastic Pion-Pion Scattering}",
    reportnumber = "CERN-TH-5555/89",
    doi = "10.1016/0370-2693(90)90109-J",
    journal = "Phys. Lett. B",
    volume = "235",
    pages = "134--140",
    year = "1990"
}

@article{Dobado:1996ps,
    author = "Dobado, A. and Pelaez, J. R.",
    title = "{The Inverse amplitude method in chiral perturbation theory}",
    eprint = "hep-ph/9604416",
    archiveprefix = "arXiv",
    reportnumber = "LBL-38645, UCM-FT-3-96",
    doi = "10.1103/PhysRevD.56.3057",
    journal = "Phys. Rev. D",
    volume = "56",
    pages = "3057--3073",
    year = "1997"
}

@book{Eden:1966dnq,
    author = "Eden, Richard John and Landshoff, Peter V. and Olive, David I. and Polkinghorne, John Charlton",
    title = "{The analytic S-matrix}",
    isbn = "978-0-521-04869-9",
    publisher = "Cambridge Univ. Press",
    address = "Cambridge",
    year = "1966"
}

@article{Estabrooks:1974vu,
    author = "Estabrooks, P. and Martin, Alan D.",
    title = "{$\pi \pi$ Phase Shift Analysis Below the $K \bar K$ Threshold}",
    reportnumber = "Print-74-0966 (DURHAM)",
    doi = "10.1016/0550-3213(74)90488-X",
    journal = "Nucl. Phys. B",
    volume = "79",
    pages = "301--316",
    year = "1974"
}

@article{Etkin:1981sg,
    author = "Etkin, A. and others",
    title = "{Amplitude Analysis of the $K^0_S K^0_S$ System Produced in the Reaction $\pi^- p \to K^0_S K^0_S n$ at 23 GeV$/c$}",
    reportnumber = "BNL-30083-R",
    doi = "10.1103/PhysRevD.25.1786",
    journal = "Phys. Rev. D",
    volume = "25",
    pages = "1786",
    year = "1982"
}

@article{Fornes:2025ldd,
    author = "Forn{\'e}s, Marta Campos and Mathieu, Vincent",
    title = "{Barrelet Zeros Extraction in Pion-Nucleon Scattering}",
    doi = "10.22323/1.465.0047",
    journal = "PoS",
    volume = "QNP2024",
    pages = "047",
    year = "2025"
}

@article{GarciaMartin:2011cn,
    author = "Garcia-Martin, R. and Kaminski, R. and Pelaez, J. R. and Ruiz de Elvira, J. and Yndurain, F. J.",
    title = "{The Pion-pion scattering amplitude. IV: Improved analysis with once subtracted Roy-like equations up to 1100 MeV}",
    eprint = "1102.2183",
    archiveprefix = "arXiv",
    primaryclass = "hep-ph",
    doi = "10.1103/PhysRevD.83.074004",
    journal = "Phys. Rev. D",
    volume = "83",
    pages = "074004",
    year = "2011"
}

@article{GarciaMartin:2011jx,
    author = "Garcia-Martin, R. and Kaminski, R. and Pelaez, J. R. and Ruiz de Elvira, J.",
    title = "{Precise determination of the $f_0(600)$ and $f_0(980)$ pole parameters from a dispersive data analysis}",
    eprint = "1107.1635",
    archiveprefix = "arXiv",
    primaryclass = "hep-ph",
    doi = "10.1103/PhysRevLett.107.072001",
    journal = "Phys. Rev. Lett.",
    volume = "107",
    pages = "072001",
    year = "2011"
}

@article{Garrote:2022uub,
    author = "Garrote, R. Alvarez and Cuervo, J. and Magalh{\~a}es, P. C. and Pel{\'a}ez, J. R.",
    title = "{Dispersive $\pi\pi \to KK$ Amplitude and Giant $CP$ Violation in $B$ to Three Light-Meson Decays at LHCb}",
    eprint = "2210.08354",
    archiveprefix = "arXiv",
    primaryclass = "hep-ph",
    reportnumber = "IPARCOS-UCM-23-033",
    doi = "10.1103/PhysRevLett.130.201901",
    journal = "Phys. Rev. Lett.",
    volume = "130",
    number = "20",
    pages = "201901",
    year = "2023"
}

@article{Gasser:1983yg,
    author = "Gasser, J. and Leutwyler, H.",
    title = "{Chiral Perturbation Theory to One Loop}",
    reportnumber = "CERN-TH-3689",
    doi = "10.1016/0003-4916(84)90242-2",
    journal = "Annals Phys.",
    volume = "158",
    pages = "142",
    year = "1984"
}

@article{Gasser:1984gg,
    author = "Gasser, J. and Leutwyler, H.",
    title = "{Chiral Perturbation Theory: Expansions in the Mass of the Strange Quark}",
    reportnumber = "CERN-TH-3798",
    doi = "10.1016/0550-3213(85)90492-4",
    journal = "Nucl. Phys. B",
    volume = "250",
    pages = "465--516",
    year = "1985"
}

@article{GomezNicola:2001as,
    author = "Gomez Nicola, A. and Pelaez, J. R.",
    title = "{Meson meson scattering within one loop chiral perturbation theory and its unitarization}",
    eprint = "hep-ph/0109056",
    archiveprefix = "arXiv",
    doi = "10.1103/PhysRevD.65.054009",
    journal = "Phys. Rev. D",
    volume = "65",
    pages = "054009",
    year = "2002"
}

@article{Grayer:1974cr,
    author = "Grayer, G. and others",
    title = "{High Statistics Study of the Reaction $\pi^- p \to \pi^- \pi^+ n$: Apparatus, Method of Analysis, and General Features of Results at 17-GeV/c}",
    reportnumber = "Print-74-0701 (CERN)",
    doi = "10.1016/0550-3213(74)90545-8",
    journal = "Nucl. Phys. B",
    volume = "75",
    pages = "189--245",
    year = "1974"
}

@article{Guerrieri:2018uew,
    author = "Guerrieri, Andrea L. and Penedones, Joao and Vieira, Pedro",
    title = "{Bootstrapping QCD Using Pion Scattering Amplitudes}",
    eprint = "1810.12849",
    archiveprefix = "arXiv",
    primaryclass = "hep-th",
    doi = "10.1103/PhysRevLett.122.241604",
    journal = "Phys. Rev. Lett.",
    volume = "122",
    number = "24",
    pages = "241604",
    year = "2019"
}

@article{Guerrieri:2024jkn,
    author = "Guerrieri, Andrea and H{\"a}ring, Kelian and Su, Ning",
    title = "{From data to the analytic S-matrix: A Bootstrap fit of the pion scattering amplitude}",
    eprint = "2410.23333",
    archiveprefix = "arXiv",
    primaryclass = "hep-th",
    reportnumber = "CERN-TH-2024-177, CALT-TH-2024-038",
    doi = "10.21468/SciPostPhys.20.2.034",
    journal = "SciPost Phys.",
    volume = "20",
    number = "2",
    pages = "034",
    year = "2026"
}

@article{Gumus:2024lmj,
    author = "Gumus, Mehmet Asim and Leflot, Damien and Tourkine, Piotr and Zhiboedov, Alexander",
    title = "{The S-matrix bootstrap with neural optimizers. Part I. Zero double discontinuity}",
    eprint = "2412.09610",
    archiveprefix = "arXiv",
    primaryclass = "hep-th",
    reportnumber = "LAPTh-061/24; CERN-TH-2024-209",
    doi = "10.1007/JHEP07(2025)210",
    journal = "JHEP",
    volume = "07",
    pages = "210",
    year = "2025"
}

@article{Hanhart:2008mx,
    author = "Hanhart, C. and Pelaez, J. R. and Rios, G.",
    title = "{Quark mass dependence of the $\rho$ and $\sigma$ from dispersion relations and Chiral Perturbation Theory}",
    eprint = "0801.2871",
    archiveprefix = "arXiv",
    primaryclass = "hep-ph",
    reportnumber = "FZJ-IKP-TH-2008-01",
    doi = "10.1103/PhysRevLett.100.152001",
    journal = "Phys. Rev. Lett.",
    volume = "100",
    pages = "152001",
    year = "2008"
}

@article{Hanhart:2014ssa,
    author = "Hanhart, C. and Pelaez, J. R. and Rios, G.",
    title = "{Remarks on pole trajectories for resonances}",
    eprint = "1407.7452",
    archiveprefix = "arXiv",
    primaryclass = "hep-ph",
    doi = "10.1016/j.physletb.2014.11.011",
    journal = "Phys. Lett. B",
    volume = "739",
    pages = "375--382",
    year = "2014"
}

@article{Hoferichter:2023mgy,
    author = "Hoferichter, Martin and Ruiz de Elvira, Jacobo and Kubis, Bastian and Mei{\ss}ner, Ulf-G.",
    title = "{Nucleon resonance parameters from Roy-Steiner equations}",
    eprint = "2312.15015",
    archiveprefix = "arXiv",
    primaryclass = "hep-ph",
    reportnumber = "IPARCOS-UCM-23-139",
    doi = "10.1016/j.physletb.2024.138698",
    journal = "Phys. Lett. B",
    volume = "853",
    pages = "138698",
    year = "2024"
}

@article{Hyams:1973zf,
    author = "Hyams, B. and others",
    title = "{$\pi\pi$ Phase Shift Analysis from 600 MeV to 1900 MeV}",
    doi = "10.1016/0550-3213(73)90618-4",
    journal = "Nucl. Phys. B",
    volume = "64",
    pages = "134--162",
    year = "1973"
}

@article{Hyams:1975mc,
    author = "Hyams, B. and others",
    title = "{A Study of All the $\pi \pi$ Phase Shift Solutions in the Mass Region 1.0 GeV to 1.8 GeV from $\pi^- p \to \pi^- \pi^+ n$ at 17.2 GeV}",
    reportnumber = "PRINT-75-0527 (CERN)",
    doi = "10.1016/0550-3213(75)90616-1",
    journal = "Nucl. Phys. B",
    volume = "100",
    pages = "205--224",
    year = "1975"
}

@article{Jaffe:2004ph,
    author = "Jaffe, R. L.",
    editor = "Kunihiro, Teiji and Onogi, Tetsuya and Abuki, H. and Takahashi, Toru T.",
    title = "{Exotica}",
    eprint = "hep-ph/0409065",
    archiveprefix = "arXiv",
    reportnumber = "MIT-CTP-3538",
    doi = "10.1016/j.physrep.2004.11.005",
    journal = "Phys. Rept.",
    volume = "409",
    pages = "1--45",
    year = "2005"
}

@article{Johnson:2024omq,
    author = "Johnson, Daniel and Polyakov, Ivan and Skwarnicki, Tomasz and Wang, Mengzhen",
    title = "{Exotic Hadrons at LHCb}",
    eprint = "2403.04051",
    archiveprefix = "arXiv",
    primaryclass = "hep-ex",
    doi = "10.1146/annurev-nucl-102422-040628",
    journal = "Ann. Rev. Nucl. Part. Sci.",
    volume = "74",
    number = "1",
    pages = "583--612",
    year = "2024"
}

@article{JointPhysicsAnalysisCenter:2023gku,
    author = "Smith, W. A. and others",
    collaboration = "Joint Physics Analysis Center",
    title = "{Ambiguities in partial wave analysis of two spinless meson photoproduction}",
    eprint = "2306.17779",
    archiveprefix = "arXiv",
    primaryclass = "hep-ph",
    reportnumber = "JLAB-THY-23-3873",
    doi = "10.1103/PhysRevD.108.076001",
    journal = "Phys. Rev. D",
    volume = "108",
    number = "7",
    pages = "076001",
    year = "2023"
}

@article{JPAC:2018zyd,
    author = "Rodas, A. and others",
    collaboration = "JPAC",
    title = "{Determination of the pole position of the lightest hybrid meson candidate}",
    eprint = "1810.04171",
    archiveprefix = "arXiv",
    primaryclass = "hep-ph",
    reportnumber = "JLAB-THY-18-2839",
    doi = "10.1103/PhysRevLett.122.042002",
    journal = "Phys. Rev. Lett.",
    volume = "122",
    number = "4",
    pages = "042002",
    year = "2019"
}

@article{JPAC:2021rxu,
    author = "Albaladejo, Miguel and others",
    collaboration = "JPAC",
    title = "{Novel approaches in hadron spectroscopy}",
    eprint = "2112.13436",
    archiveprefix = "arXiv",
    primaryclass = "hep-ph",
    reportnumber = "LA-UR-21-31664, JLAB-THY-22-3459",
    doi = "10.1016/j.ppnp.2022.103981",
    journal = "Prog. Part. Nucl. Phys.",
    volume = "127",
    pages = "103981",
    year = "2022"
}

@article{Kaminski:2002pe,
    author = "Kaminski, R. and Lesniak, L. and Loiseau, B.",
    title = "{Elimination of ambiguities in $\pi \pi$ phase shifts using crossing symmetry}",
    eprint = "hep-ph/0210334",
    archiveprefix = "arXiv",
    doi = "10.1016/S0370-2693(02)03021-6",
    journal = "Phys. Lett. B",
    volume = "551",
    pages = "241--248",
    year = "2003"
}

@article{LHCb:2019jta,
    author = "Aaij, Roel and others",
    collaboration = "LHCb",
    title = "{Observation of Several Sources of $CP$ Violation in $B^+ \to \pi^+ \pi^+ \pi^-$ Decays}",
    eprint = "1909.05211",
    archiveprefix = "arXiv",
    primaryclass = "hep-ex",
    reportnumber = "LHCb-PAPER-2019-018, CERN-EP-2019-156",
    doi = "10.1103/PhysRevLett.124.031801",
    journal = "Phys. Rev. Lett.",
    volume = "124",
    number = "3",
    pages = "031801",
    year = "2020"
}

@article{Londergan:2013dza,
    author = "Londergan, J. T. and Nebreda, J. and Pelaez, J. R. and Szczepaniak, A.",
    title = "{Identification of non-ordinary mesons from the dispersive connection between their poles and their Regge trajectories: The $f_0(500)$ resonance}",
    eprint = "1311.7552",
    archiveprefix = "arXiv",
    primaryclass = "hep-ph",
    reportnumber = "JLAB-THY-14-1837",
    doi = "10.1016/j.physletb.2013.12.061",
    journal = "Phys. Lett. B",
    volume = "729",
    pages = "9--14",
    year = "2014"
}

@article{Masjuan:2014psa,
    author = "Masjuan, Pere and Ruiz de Elvira, Jacobo and Sanz-Cillero, Juan Jos{\'e}",
    title = "{Precise determination of resonance pole parameters through Pad{\'e} approximants}",
    eprint = "1410.2397",
    archiveprefix = "arXiv",
    primaryclass = "hep-ph",
    reportnumber = "MITP-14-070, IFT-UAM-CSIC-14-100, FTUAM-14-38",
    doi = "10.1103/PhysRevD.90.097901",
    journal = "Phys. Rev. D",
    volume = "90",
    number = "9",
    pages = "097901",
    year = "2014"
}

@article{Morgan:1990ct,
    author = "Morgan, D. and Pennington, M. R.",
    title = "{$f_0 (S*)$: Molecule or quark state?}",
    reportnumber = "RAL-90-090",
    doi = "10.1016/0370-2693(91)91115-C",
    journal = "Phys. Lett. B",
    volume = "258",
    pages = "444--450",
    year = "1991",
    note = "[Erratum: Phys.Lett.B 269, 477 (1991)]"
}

@article{Moussallam:2011zg,
    author = "Moussallam, B.",
    title = "{Couplings of light $I=0$ scalar mesons to simple operators in the complex plane}",
    eprint = "1110.6074",
    archiveprefix = "arXiv",
    primaryclass = "hep-ph",
    doi = "10.1140/epjc/s10052-011-1814-z",
    journal = "Eur. Phys. J. C",
    volume = "71",
    pages = "1814",
    year = "2011"
}

@article{Ng:2021ibr,
    author = "Ng, L. and Bibrzycki, L. and Nys, J. and Fernandez-Ramirez, C. and Pilloni, A. and Mathieu, V. and Rasmusson, A. J. and Szczepaniak, A. P.",
    collaboration = "Joint Physics Analysis Center, JPAC",
    title = "{Deep learning exotic hadrons}",
    eprint = "2110.13742",
    archiveprefix = "arXiv",
    primaryclass = "hep-ph",
    reportnumber = "JLAB-THY-21-3518",
    doi = "10.1103/PhysRevD.105.L091501",
    journal = "Phys. Rev. D",
    volume = "105",
    number = "9",
    pages = "L091501",
    year = "2022"
}

@article{Nieves:1999bx,
    author = "Nieves, Juan and Ruiz Arriola, Enrique",
    title = "{Bethe-Salpeter approach for unitarized chiral perturbation theory}",
    eprint = "hep-ph/9907469",
    archiveprefix = "arXiv",
    reportnumber = "UG-DFM-2-99",
    doi = "10.1016/S0375-9474(00)00321-3",
    journal = "Nucl. Phys. A",
    volume = "679",
    pages = "57--117",
    year = "2000"
}

@article{NNPDF:2021njg,
    author = "Ball, Richard D. and others",
    collaboration = "NNPDF",
    title = "{The path to proton structure at 1{\%} accuracy}",
    eprint = "2109.02653",
    archiveprefix = "arXiv",
    primaryclass = "hep-ph",
    reportnumber = "Edinburgh 2021/12, Nikhef-2021-013, TIF-UNIMI-2021-11",
    doi = "10.1140/epjc/s10052-022-10328-7",
    journal = "Eur. Phys. J. C",
    volume = "82",
    number = "5",
    pages = "428",
    year = "2022"
}

@article{Ochs:2013gi,
    author = "Ochs, Wolfgang",
    title = "{The Status of Glueballs}",
    eprint = "1301.5183",
    archiveprefix = "arXiv",
    primaryclass = "hep-ph",
    reportnumber = "MPP-2013-44",
    doi = "10.1088/0954-3899/40/4/043001",
    journal = "J. Phys. G",
    volume = "40",
    pages = "043001",
    year = "2013"
}

@article{Oller:1997ti,
    author = "Oller, J. A. and Oset, E.",
    title = "{Chiral symmetry amplitudes in the S wave isoscalar and isovector channels and the $\sigma$, $f_0$(980), $a_0$(980) scalar mesons}",
    eprint = "hep-ph/9702314",
    archiveprefix = "arXiv",
    doi = "10.1016/S0375-9474(97)00160-7",
    journal = "Nucl. Phys. A",
    volume = "620",
    pages = "438--456",
    year = "1997",
    note = "[Erratum: Nucl.Phys.A 652, 407--409 (1999)]"
}

@article{Oller:1998hw,
    author = "Oller, J. A. and Oset, E. and Pelaez, J. R.",
    title = "{Meson meson interaction in a nonperturbative chiral approach}",
    eprint = "hep-ph/9804209",
    archiveprefix = "arXiv",
    reportnumber = "SLAC-PUB-7787",
    doi = "10.1103/PhysRevD.59.074001",
    journal = "Phys. Rev. D",
    volume = "59",
    pages = "074001",
    year = "1999",
    note = "[Erratum: Phys.Rev.D 60, 099906 (1999), Erratum: Phys.Rev.D 75, 099903 (2007)]"
}

@article{Oller:1998zr,
    author = "Oller, J. A. and Oset, E.",
    title = "{$N/D$ description of two meson amplitudes and chiral symmetry}",
    eprint = "hep-ph/9809337",
    archiveprefix = "arXiv",
    doi = "10.1103/PhysRevD.60.074023",
    journal = "Phys. Rev. D",
    volume = "60",
    pages = "074023",
    year = "1999"
}

@article{ParticleDataGroup:2024cfk,
    author = "Navas, S. and others",
    collaboration = "Particle Data Group",
    title = "{Review of particle physics}",
    doi = "10.1103/PhysRevD.110.030001",
    journal = "Phys. Rev. D",
    volume = "110",
    number = "3",
    pages = "030001",
    year = "2024"
}

@article{Pelaez:2004xp,
    author = "Pelaez, J. R.",
    title = "{Light scalars as tetraquarks or two-meson states from large $N_c$ and unitarized chiral perturbation theory}",
    eprint = "hep-ph/0411107",
    archiveprefix = "arXiv",
    doi = "10.1142/S0217732304016160",
    journal = "Mod. Phys. Lett. A",
    volume = "19",
    pages = "2879--2894",
    year = "2004"
}

@article{Pelaez:2015qba,
    author = "Pelaez, J. R.",
    title = "{From controversy to precision on the sigma meson: a review on the status of the non-ordinary $f_0(500)$ resonance}",
    eprint = "1510.00653",
    archiveprefix = "arXiv",
    primaryclass = "hep-ph",
    doi = "10.1016/j.physrep.2016.09.001",
    journal = "Phys. Rept.",
    volume = "658",
    pages = "1",
    year = "2016"
}

@article{Pelaez:2020gnd,
    author = "Pel{\'a}ez, Jos{\'e} R. and Rodas, Arkaitz",
    title = "{Dispersive $\pi K \to \pi K$ and $1pi1pi \to K \bar K$ amplitudes from scattering data, threshold parameters, and the lightest strange resonance $\kappa$ or $K_0^*(700)$}",
    eprint = "2010.11222",
    archiveprefix = "arXiv",
    primaryclass = "hep-ph",
    reportnumber = "JLAB-THY-20-3276",
    doi = "10.1016/j.physrep.2022.03.004",
    journal = "Phys. Rept.",
    volume = "969",
    pages = "1--126",
    year = "2022"
}

@article{Pelaez:2026fdrpoles,
    author = "Pel{\'a}ez, Jos{\'e} Ram{\'o}n and Rab{\'a}n, Pablo and de Elvira, Jacobo Ruiz",
    title = "{Dispersive determination of resonances from $\pi\pi$ scattering data}",
    eprint = "2512.09033",
    archiveprefix = "arXiv",
    primaryclass = "hep-ph",
    reportnumber = "IPARCOS-UCM-25-065",
    doi = "10.1103/fdv7-k23j",
    journal = "Phys. Rev. D",
    volume = "113",
    number = "3",
    pages = "034018",
    year = "2026"
}

@article{Perez:2015pea,
    author = "Navarro P{\'e}rez, R. and Ruiz Arriola, E. and Ruiz de Elvira, J.",
    title = "{Self-consistent statistical error analysis of $\pi\pi$ scattering}",
    eprint = "1502.03361",
    archiveprefix = "arXiv",
    primaryclass = "hep-ph",
    doi = "10.1103/PhysRevD.91.074014",
    journal = "Phys. Rev. D",
    volume = "91",
    pages = "074014",
    year = "2015"
}

@article{Protopopescu:1973sh,
    author = "Protopopescu, S. D. and Alston-Garnjost, M. and Barbaro-Galtieri, A. and Flatte, Stanley M. and Friedman, J. H. and Lasinski, T. A. and Lynch, G. R. and Rabin, M. S. and Solmitz, F. T.",
    title = "{$\pi\pi$ Partial Wave Analysis from Reactions $\pi^+ p \to \pi^+ \pi^- \Delta^{++}$ and $\pi^+ p \to K^+ K^- \Delta^{++}$ at 7.1 GeV$/c$}",
    reportnumber = "LBL-787",
    doi = "10.1103/PhysRevD.7.1279",
    journal = "Phys. Rev. D",
    volume = "7",
    pages = "1279",
    year = "1973"
}

@article{Rodas:2021tyb,
    author = "Rodas, A. and Pilloni, A. and Albaladejo, M. and Fernandez-Ramirez, C. and Mathieu, V. and Szczepaniak, A. P.",
    collaboration = "Joint Physics Analysis Center",
    title = "{Scalar and tensor resonances in $J/\psi$ radiative decays}",
    eprint = "2110.00027",
    archiveprefix = "arXiv",
    primaryclass = "hep-ph",
    doi = "10.1140/epjc/s10052-022-10014-8",
    journal = "Eur. Phys. J. C",
    volume = "82",
    number = "1",
    pages = "80",
    year = "2022"
}

@article{Rodas:2023nec,
    author = "Rodas, Arkaitz and Dudek, Jozef J. and Edwards, Robert G.",
    collaboration = "Hadron Spectrum",
    title = "{Determination of crossing-symmetric $\pi\pi$ scattering amplitudes and the quark mass evolution of the $\sigma$ constrained by lattice QCD}",
    eprint = "2304.03762",
    archiveprefix = "arXiv",
    primaryclass = "hep-lat",
    reportnumber = "JLAB-THY-23-3791",
    doi = "10.1103/PhysRevD.109.034513",
    journal = "Phys. Rev. D",
    volume = "109",
    number = "3",
    pages = "034513",
    year = "2024"
}

@article{RuizdeElvira:2010cs,
    author = "Ruiz de Elvira, J. and Pelaez, J. R. and Pennington, M. R. and Wilson, D. J.",
    title = "{Chiral Perturbation Theory, the ${1/N_c}$ expansion and Regge behaviour determine the structure of the lightest scalar meson}",
    eprint = "1009.6204",
    archiveprefix = "arXiv",
    primaryclass = "hep-ph",
    reportnumber = "DCPT-10-98, IPPP-10-49, JLAB-THY-10-1256",
    doi = "10.1103/PhysRevD.84.096006",
    journal = "Phys. Rev. D",
    volume = "84",
    pages = "096006",
    year = "2011"
}

@article{Sadasivan:2025kjj,
    author = "Sadasivan, Daniel and Cordero, Isaac and Graham, Andrew and Marsh, Cecilia and Kupcho, Daniel and Mourad, Melana and Mai, Maxim",
    title = "{Deep neural network driven simulation based inference method for pole position estimation under model misspecification}",
    eprint = "2507.18824",
    archiveprefix = "arXiv",
    primaryclass = "hep-ph",
    doi = "10.1103/ncqd-f1g2",
    journal = "Phys. Rev. D",
    volume = "114",
    number = "1",
    pages = "016002",
    year = "2026"
}

@article{Shepherd:2016dni,
    author = "Shepherd, Matthew R. and Dudek, Jozef J. and Mitchell, Ryan E.",
    title = "{Searching for the rules that govern hadron construction}",
    eprint = "1802.08131",
    archiveprefix = "arXiv",
    primaryclass = "hep-ph",
    reportnumber = "JLAB-THY-16-2286",
    doi = "10.1038/nature18011",
    journal = "Nature",
    volume = "534",
    number = "7608",
    pages = "487--493",
    year = "2016"
}

@article{Sombillo:2021yxe,
    author = "Sombillo, Denny Lane B. and Ikeda, Yoichi and Sato, Toru and Hosaka, Atsushi",
    title = "{Unveiling the pole structure of S-matrix using deep learning}",
    eprint = "2104.14182",
    archiveprefix = "arXiv",
    primaryclass = "hep-ph",
    doi = "10.31349/SuplRevMexFis.3.0308067",
    journal = "Rev. Mex. Fis. Suppl.",
    volume = "3",
    number = "3",
    pages = "0308067",
    year = "2022"
}

@article{Weinberg:1978kz,
    author = "Weinberg, Steven",
    editor = "Deser, S.",
    title = "{Phenomenological Lagrangians}",
    reportnumber = "HUTP-78-A051A",
    doi = "10.1016/0378-4371(79)90223-1",
    journal = "Physica A",
    volume = "96",
    number = "1-2",
    pages = "327--340",
    year = "1979"
}

@article{Yndurain:2007qm,
    author = "Yndurain, F. J. and Garcia-Martin, R. and Pelaez, J. R.",
    title = "{Experimental status of the $\pi \pi$ isoscalar S-wave at low energy: $f_0(600)$ pole and scattering length}",
    eprint = "hep-ph/0701025",
    archiveprefix = "arXiv",
    reportnumber = "FTUAM-06-19",
    doi = "10.1103/PhysRevD.76.074034",
    journal = "Phys. Rev. D",
    volume = "76",
    pages = "074034",
    year = "2007"
}

@InProceedings{Ilyas2022,
    title = "Datamodels: Understanding Predictions with Data and Data with Predictions",
    author = "Ilyas, Andrew and Park, Sung Min and Engstrom, Logan and Leclerc, Guillaume and Madry, Aleksander",
    booktitle = "Proceedings of the 39th International Conference on Machine Learning",
    pages = "9525--9587",
    year = "2022",
    editor = "Chaudhuri, Kamalika and Jegelka, Stefanie and Song, Le and Szepesvari, Csaba and Niu, Gang and Sabato, Sivan",
    volume = "162",
    series = "Proceedings of Machine Learning Research",
    month = "17--23 Jul",
    publisher = "PMLR",
    url = "https://proceedings.mlr.press/v162/ilyas22a.html"
}

@InProceedings{KohLiang2017,
    title = "Understanding Black-box Predictions via Influence Functions",
    author = "Pang Wei Koh and Percy Liang",
    booktitle = "Proceedings of the 34th International Conference on Machine Learning",
    pages = "1885--1894",
    year = "2017",
    editor = "Precup, Doina and Teh, Yee Whye",
    volume = "70",
    series = "Proceedings of Machine Learning Research",
    month = "06--11 Aug",
    publisher = "PMLR",
    url = "https://proceedings.mlr.press/v70/koh17a.html"
}

@article{Murtagh2014,
    title = "Ward's Hierarchical Agglomerative Clustering Method: Which Algorithms Implement Ward's Criterion?",
    volume = "31",
    doi = "10.1007/s00357-014-9161-z",
    number = "3",
    journal = "J. Classif.",
    author = "Murtagh, Fionn and Legendre, Pierre",
    year = "2014",
    month = "Oct",
    pages = "274--295"
}

@book{Newton:1982,
    title = "Scattering Theory of Waves and Particles",
    isbn = "9783642881282",
    url = "http://dx.doi.org/10.1007/978-3-642-88128-2",
    doi = "10.1007/978-3-642-88128-2",
    publisher = "Springer Berlin Heidelberg",
    author = "Newton, Roger G.",
    year = "1982"
}

@unpublished{Talts:2018sbc,
    title = "Validating Bayesian Inference Algorithms with Simulation-Based Calibration",
    author = "Sean Talts and Michael Betancourt and Daniel Simpson and Aki Vehtari and Andrew Gelman",
    year = "2020",
    eprint = "1804.06788",
    archiveprefix = "arXiv",
    primaryclass = "stat.ME",
    url = "https://arxiv.org/abs/1804.06788"
}

@article{Ward01031963,
  author = {Ward Jr., Joe H.},
  title = {Hierarchical Grouping to Optimize an Objective Function},
  journal = {J. Am. Stat. Assoc.},
  volume = {58},
  number = {301},
  pages = {236--244},
  year = {1963},
  doi = {10.1080/01621459.1963.10500845}
}

@misc{wyatt_a_smith_2026_22016314,
    doi = "10.5281/ZENODO.22016314",
    url = "https://zenodo.org/doi/10.5281/zenodo.22016314",
    author = "Smith, Wyatt A. and others",
    title = "smithwya/pipi\_SINN\_26: Production SINN ensemble for pion scattering",
    publisher = "Zenodo",
    year = "2026"
}

@book{GoldbergerWatson:1964,
    author = "Goldberger, Marvin L. and Watson, Kenneth M.",
    title = "{Collision Theory}",
    publisher = "Wiley",
    year = "1964",
   isbn={9780882753133},
  lccn={lc64017819},
  series={Structure of matter series},
  url={https://books.google.it/books?id=MR5RAAAAMAAJ},
}

@unpublished{companion,
    author = "Smith, Wyatt A. and others",
    title = "{S-matrix informed neural networks for amplitude analysis}",
    note = "Companion paper, in preparation",
    year = 2026
}

\clearpage
\newpage
\appendix
\setcounter{secnumdepth}{2}
\onecolumngrid
\section*{End matter}
\twocolumngrid

\section{Pole extraction}
\label{app:poles}
Resonance pole parameters are obtained by analytically continuing the dispersive $\pi\pi$ amplitude through the Roy equations. In terms of the partial-wave S matrix, the first two Riemann sheets are related by,
\begin{align}
\mathrm{S}_\ell^I(s) &= 1+2i\,\rho(s)\,t_\ell^I(s), \nonumber \\
\mathrm{S}_{\ell,\mathrm{II}}^I(s) &= \frac{1}{\mathrm{S}_\ell^I(s)}, \qquad
t_{\ell,\mathrm{II}}^I(s)=\frac{t_\ell^I(s)}{\mathrm{S}_\ell^I(s)},
\end{align}
where $s$ is understood to be complex. The corresponding coupling is extracted from the residue of the second-sheet amplitude using,
\begin{align}
g^2=-16\pi(2\ell+1)\lim_{s\rightarrow s_{\text{pole}}}
\frac{(s-s_{\text{pole}})\,t^I_{\ell,\mathrm{II}}(s)}{(2k)^{2\ell}}~.
\label{eq:endmatter_coupling}
\end{align}

\section{Architecture dependence}
\label{app:arch}
\begin{figure*}[t]
\begin{tabular}{c}
\includegraphics[width=\linewidth]{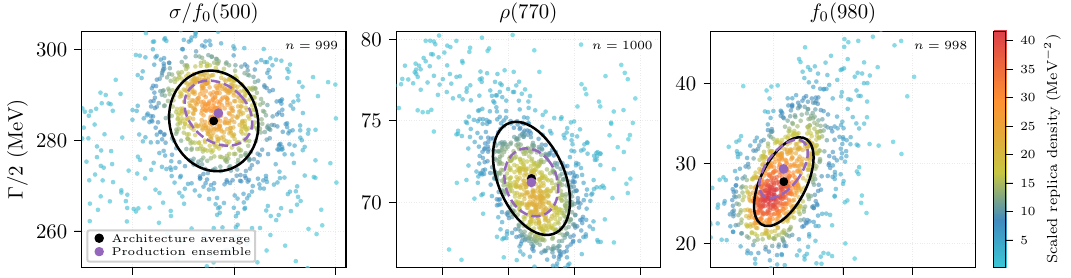} \\
\includegraphics[width=\linewidth]{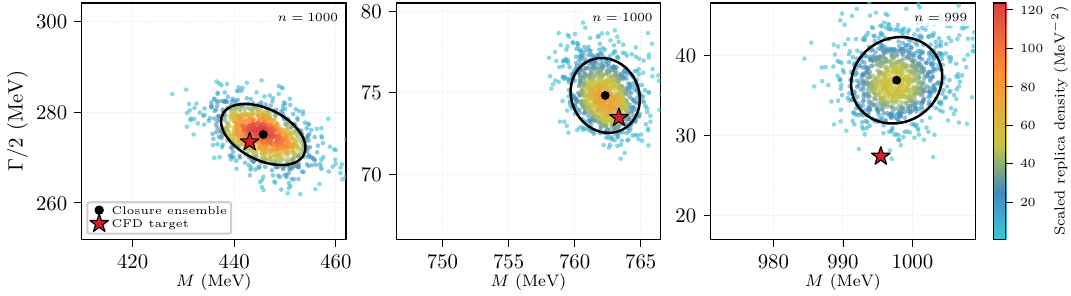}
\end{tabular}
\caption{Top row: Architecture average ($1{,}000$ networks, $100$ architectures) pole positions compared to production poles. Points are the architecture ensemble, colored by local density; the solid ellipse and dot are its median and $68\%$ uncertainty band. The dashed ellipse and dot correspond to the production pole locations. Bottom row: Closure test, pole positions in the complex plane for the $1{,}000$-replica closure ensemble. Stars mark the CFD target obtained by applying our dispersive pole search to the generating amplitude. The target is recovered for all resonance parameters except the decay width of the $f_0(980)$, which can be associated to the explicit inclusion of the $\bar KK$ threshold in the CFD analysis.}
\label{fig:pole_stability}
\end{figure*}

We test for hidden model dependence set by the choice of network with a $500$-trial sweep over $21$ hyperparameter dimensions. We retain the $100$ best architectures and train $10$ replicas for each, giving an ensemble of $1{,}000$ networks. The top row in~\cref{fig:pole_stability} compares the poles extracted from this ensemble with the production replica ensemble. The medians of every pole parameter, scattering length, and Adler zero are statistically indistinguishable from the production ensemble. Both ensembles are built from data replicas, so the architecture ensemble carries statistical uncertainties in addition to the extra spread from varying the architecture. \Cref{tab:robustness} lists the median and the central $68\%$ interval of every extracted observable. Every uncertainty quoted in this Letter is that interval taken over the production ensemble, written $\delta x$ for an observable $x$. The dashed contours of~\cref{fig:scatt,fig:adler,fig:poles} add the spread of the $100$ per-architecture medians to it in quadrature, to show the approximate systematic errors in our production uncertainty bands.

\begin{table}[!htb]
\caption{Average of medians and central $68\%$ intervals of every extracted observable over the $1{,}000$-network architecture ensemble compared to the production values.}
\label{tab:robustness}
\centering
\footnotesize
\setlength{\tabcolsep}{3pt}
\begin{tabular}{l c c}
\toprule
Observable & Production & Architecture average\\
\midrule
$M_{\sigma}$ (MeV)             & $436.9^{+6.9}_{-6.5}$    & $436.0^{+8.6}_{-8.9}$ \\[2pt]
$\Gamma_{\sigma}/2$ (MeV)      & $285.9^{+6.5}_{-7.9}$    & $284.2^{+9.1}_{-13.1}$ \\[2pt]
$|g_{\sigma\pi\pi}|$ (GeV)     & $3.33^{+0.06}_{-0.06}$   & $3.32^{+0.09}_{-0.11}$ \\
\midrule
$M_{\rho(770)}$ (MeV)          & $756.7^{+2.1}_{-2.0}$    & $756.7^{+2.7}_{-3.2}$ \\[2pt]
$\Gamma_{\rho(770)}/2$ (MeV)   & $71.2^{+2.3}_{-2.0}$     & $71.5^{+4.2}_{-2.8}$ \\[2pt]
$|g_{\rho\pi\pi}|$             & $6.00^{+0.08}_{-0.07}$   & $6.01^{+0.14}_{-0.09}$ \\
\midrule
$M_{f_0(980)}$ (MeV)           & $981.5^{+3.7}_{-3.2}$    & $981.5^{+4.8}_{-3.8}$ \\[2pt]
$\Gamma_{f_0(980)}/2$ (MeV)    & $29.3^{+4.2}_{-3.2}$     & $27.7^{+6.5}_{-4.6}$ \\[2pt]
$|g_{f_0\pi\pi}|$ (GeV)        & $1.87^{+0.15}_{-0.11}$   & $1.82^{+0.24}_{-0.15}$ \\
\midrule
$a_0^0$ ($m_\pi^{-1}$)         & $0.222^{+0.014}_{-0.014}$  & $0.222^{+0.013}_{-0.017}$ \\[2pt]
$a_0^2$ ($m_\pi^{-1}$)         & $-0.041^{+0.005}_{-0.006}$ & $-0.041^{+0.005}_{-0.009}$ \\
\midrule
$S_0$ Adler zero ($m_\pi^{2}$) & $0.34^{+0.26}_{-0.22}$   & $0.38^{+0.39}_{-0.22}$ \\[2pt]
$S_2$ Adler zero ($m_\pi^{2}$) & $2.09^{+0.20}_{-0.21}$   & $2.02^{+0.16}_{-0.28}$ \\
\bottomrule
\end{tabular}
\end{table}

\section{Role of the Roy constraints}
\label{app:ablation}
To assess the effects of the Roy equations, we retrain a $1{,}000$-replica ablated ensemble with $\lambda_{\text{Roy}}=0$, holding the architecture, curriculum, and data fixed. \Cref{tab:ablation} collects the comparison between this ensemble and our production ensemble.

The ablated ensemble and production ensemble are indistinguishable based off data satisfaction. All observables extracted via analytic continuation through the Roy equations evaluated on the ablated ensemble are significantly shifted and exhibit much larger uncertainty bands compared to the production ensemble extractions. Despite this, extraction of continued quantities is relatively stable, except notably for the location of the $f_0(980)$ pole which is recoverable in only $241$ of the $1{,}000$ unconstrained replicas, contrasted to the production ensemble in which every SINN was able to reliably produce a pole. We conclude that imposing Roy constraints leaves the real-axis data description unchanged while substantially stabilizing the subthreshold and off-axis observables.

\begin{table}[!htb]
\caption{Production versus ablated SINN ensembles, $10^4$ and $10^3$ replicas. Intervals are central $68\%$ uncerainties. See~\cite{companion} for definition of the data ($\mathcal{L}_{\text{data}}$) and Roy ($\mathcal{L}_{\text{Roy}}$) losses.}
\label{tab:ablation}
\centering
\small
\setlength{\tabcolsep}{4pt}
\begin{tabular}{l c c}
\toprule
 & Production & Ablation \\
\midrule
$\mathcal{L}_{\text{data}}$   & $2.93^{+0.21}_{-0.19}$ & $2.93^{+0.23}_{-0.21}$ \\[2pt]
$\mathcal{L}_{\text{Roy}}$ & $0.067^{+0.013}_{-0.011}$ & $0.74^{+0.30}_{-0.41}$ \\
\midrule
$M_{\sigma}$ (MeV)          & $436.9^{+6.9}_{-6.5}$   & $400.0^{+28.7}_{-13.4}$ \\[2pt]
$\Gamma_{\sigma}/2$ (MeV)     & $285.9^{+6.5}_{-7.9}$ & $199.3^{+39.9}_{-15.6}$ \\[2pt]
$M_{\rho(770)}$ (MeV)       & $756.7^{+2.1}_{-2.0}$   & $773.6^{+9.5}_{-13.7}$ \\[2pt]
$\Gamma_{\rho(770)}/2$ (MeV)  & $71.2^{+2.3}_{-2.0}$   & $79.3^{+4.8}_{-6.5}$ \\[2pt]
$M_{f_0(980)}$ (MeV)        & $981.5^{+3.7}_{-3.2}$   & $964.7^{+22.0}_{-6.0}$ \\[2pt]
$\Gamma_{f_0(980)}/2$ (MeV)   & $29.3^{+4.2}_{-3.2}$    & $19.2^{+8.1}_{-6.4}$ \\
\midrule
$S_0$ Adler zero ($m_\pi^{2}$) & $0.34^{+0.26}_{-0.22}$ & $2.06^{+0.38}_{-0.71}$ \\[2pt]
$S_2$ Adler zero ($m_\pi^{2}$) & $2.09^{+0.20}_{-0.21}$ & $0.81^{+0.66}_{-0.31}$ \\
\bottomrule
\end{tabular}
\end{table}

\begin{table}[!htb]
\caption{Closure-test results for the $1{,}000$-replica CFD ensemble. Pole masses and half widths are in $\mathrm{MeV}$.}
\label{tab:closure}
\centering
\small
\begin{tabular}{l r r r}
\toprule
Observable &~~CFD Target~~&~~SINN Closure~~&~~$z$ \\
\midrule
$M_{\sigma}$                  & $443.0$   & $445.7^{+5.2}_{-5.2}$   & $+0.52$ \\[2pt]
$\Gamma_{\sigma}/2$           & $273.4$   & $275.1^{+4.1}_{-4.4}$   & $+0.39$ \\[2pt]
$M_{\rho(770)}$               & $763.3$   & $762.3^{+1.4}_{-1.5}$   & $-0.70$ \\[2pt]
$\Gamma_{\rho(770)}/2$        & $73.5$    & $74.9^{+1.6}_{-1.3}$    & $+0.96$ \\[2pt]
$M_{f_0(980)}$                & $995.4$   & $997.7^{+4.0}_{-4.0}$   & $+0.57$ \\[2pt]
$\Gamma_{f_0(980)}/2$         & $27.4$    & $36.9^{+3.7}_{-3.0}$    & $+2.85$ \\
\bottomrule
\end{tabular}
\end{table}

\section{Closure test}
\label{app:closure_em}
Following the NNPDF validation strategy~\cite{NNPDF:2021njg,DelDebbio:2021whr} and simulation-based calibration~\cite{Talts:2018sbc}, we generate pseudo-data from a known parametrization of $\pi\pi$ scattering, the Constrained Fit to Data (CFD) parametrization~\cite{GarciaMartin:2011cn}, at the exact kinematics of the selected dataset with preserved uncertainties, then train a SINN ensemble for analysis. Because the generating amplitude is known, any displacement of the reconstructed ensemble away from it measures the bias introduced by the SINN representation of the scattering amplitude, and our training strategy. This effectively decouples data selection from the training. For an observable with closure-ensemble median $\hat{x}$ and target value \mbox{$x_{\rm target}$} we define the pull \mbox{$z\equiv(\hat{x}-x_{\rm target})/\delta x$}, with $\delta x$ taken over the closure ensemble itself, so $|z|\le1$ means the ensemble median is compatible with the target. All target values are re-calculated from our own implementation of the Roy equations.

All three pole masses are recovered near the ensemble centers, as reported in \cref{tab:closure}, and the corresponding distributions are shown in the bottom row in~\cref{fig:pole_stability}. The one discrepancy is the $f_0(980)$ width, recovered at \mbox{$36.9^{+3.7}_{-3.0}~\mathrm{MeV}$} against a target of \mbox{$27.4~\mathrm{MeV}$}, resulting in $z=+2.85$. The CFD $S_0$ phase carries an explicit $K\bar K$ cusp, while our representation retains the $K\bar K$ inelasticity opening but inserts no phase cusp. The smooth representation still recovers the $f_0(980)$ mass and the real-axis line shape, and the discrepancy appears only after continuation into the complex plane. The scattering lengths of the same ensemble are recovered without bias.

\end{document}